\documentclass[twocolumn,astrosymb,tighten,times,twocolappendix]{aastex701}

\usepackage{amsmath}
\usepackage{xcolor}
\usepackage[normalem]{ulem}
\usepackage{times}

\usepackage[
  protrusion=true,
  expansion=true,
  tracking=true,
  kerning=true,
  spacing=true
]{microtype}

\shorttitle{Dipole Polarizability of Finite Nuclei as a Probe of Neutron Stars}
\shortauthors{KOLIOGIANNIS ET AL.}
\received{October 23, 2025}
\revised{November 25, 2025}
\accepted{November 28, 2025}
\submitjournal{ApJL}
\begin{document}

\title{Dipole Polarizability of Finite Nuclei as a Probe of Neutron Stars}

\correspondingauthor{P.S. Koliogiannis}

\author[orcid=0000-0001-9326-7481,gname='Polychronis',sname='Koliogiannis']{P.S. Koliogiannis}
\affiliation{Department of Physics, Faculty of Science, University of Zagreb, Bijeni\v cka cesta 32, 10000 Zagreb, Croatia}
\affiliation{Department of Theoretical Physics, Aristotle University of Thessaloniki, 54124 Thessaloniki, Greece}
\email[show]{pkoliogi@phy.hr}

\author[orcid=0000-0002-2892-3208,gname='Esra',sname='Y{\"{u}}ksel']{E. Y{\"{u}}ksel} 
\affiliation{School of Mathematics and Physics, University of Surrey, Guildford, Surrey, GU2 7XH, United Kingdom}
\email{e.yuksel@surrey.ac.uk}

\author[orcid=0000-0002-1794-2817,gname='Tanmoy',sname='Ghosh']{T. Ghosh}
\affiliation{Department of Physics, Faculty of Science, University of Zagreb, Bijeni\v cka cesta 32, 10000 Zagreb, Croatia}
\email{tghosh@phy.hr}  

\author[orcid=0000-0002-6673-6622,gname='Nils',sname='Paar']{N. Paar}
\affiliation{Department of Physics, Faculty of Science, University of Zagreb, Bijeni\v cka cesta 32, 10000 Zagreb, Croatia}
\email{npaar@phy.hr}

\begin{abstract}

Nuclear ground state and collective excitation properties provide a means to probe the nuclear matter equation of state and establish connections between observables in finite nuclei and neutron stars. Specifically, the electric dipole polarizability, measured with high precision in various neutron-rich nuclei, serves as a robust constraint on the density dependence of the symmetry energy. In this Letter, we employ a class of relativistic energy density functionals in a twofold process: first, to link the electric dipole polarizability from recent experiments to the slope of the symmetry energy, and second, to translate this information into constraints on the tidal deformability and radii of neutron stars, in connection with multimessenger astrophysical observations from pulsars and binary neutron stars. We provide compelling evidence that the electric dipole polarizability represents a key nuclear observable to probe the neutron star properties. By significantly reducing the uncertainties in the mass-radius plane, our findings also align with recent multimessenger observations.

\end{abstract}



\section{Introduction} 
The equation of state (EOS) of neutron-rich matter governs the structure and properties of neutron stars, shaping key macroscopic characteristics such as mass, radius, and tidal deformability~\citep{Lattimer-2004,BURGIO2021103879,Lattimer2021}. Beyond its astrophysical significance, the EOS is directly linked to nuclear structure through the density dependence of the symmetry energy~\citep{ROCAMAZA201896}. In particular, the slope of the symmetry energy $L$ at saturation density $\rho_{0}$ plays a crucial role by determining the stiffness of the EOS, which, in turn, influences the neutron star's internal composition and mass-radius relation, as well as the dynamics of neutron star mergers~\citep{PhysRevLett.120.172702,Lattimer2021}. Understanding the behavior of the EOS across density scales requires an interdisciplinary approach, combining nuclear physics, gravitational-wave astronomy, high-energy astrophysics, and multimessenger observations~\citep{Goriely2010,Steiner_2010,Paar2014,Zhang2015,Ozel2016,Watts2016,Oertel2017,Abbott_2018,Tan2020,foundations1020017,Patra_2023,Imam_2024,Tsang2024}.\\
\indent Experimental data on finite nuclei provide valuable constraints on the nuclear EOS. For instance, the neutron-skin thickness has been shown to directly correlate with the $L$ parameter ~\citep{PhysRevLett.85.5296,PhysRevC.64.027302,PhysRevLett.102.122502, PhysRevC.81.041301,PhysRevLett.120.172702,Lattimer2021,PhysRevLett.126.172503}. Notably,~\citet{PhysRevLett.120.172702} and~\citet{PhysRevLett.126.172503} highlight the strong connection between neutron-skin thickness in neutron-rich nuclei and neutron star properties, reinforcing the role of nuclear structure experiments in constraining the EOS at supranuclear densities. However, precise determination of the neutron-skin thickness remains challenging even in the most recent studies using parity-violating electron scattering on $^{48}$Ca (CREX;~\citealt{Adhikari2022}) and $^{208}$Pb (PREX-2;~\citealt{Adhikari2021}) 
due to significant experimental uncertainties in measuring neutron radii in nuclei. In addition, CREX and PREX-2 data indicate tension with global energy density functional (EDF) theories~\citep{PhysRevLett.127.232501,RocaMaza2022}, yielding inconsistent results for the symmetry energy parameters and neutron-skin thickness~\citep{Yuksel2023}. Specifically, CREX indicates lower values for these quantities and a soft EOS, while PREX-2 results lead to significantly larger values and a stiff EOS, both deviating from predictions of other studies~\citep{Rokamaza2015,ROCAMAZA201896,Yuksel2023}.
Moreover,~\citet{PhysRevLett.134.192501} reported that nonperturbative QED corrections to the parity-violating asymmetry increase the inferred neutron-skin thicknesses of $^{48}$Ca and $^{208}$Pb, implying a higher pressure of neutron matter around saturation density. While these corrections suggest an adequate description for $^{48}$Ca, they still face difficulties in reproducing the large neutron skin inferred for $^{208}$Pb, thereby underscoring the remaining tension between theoretical predictions and experimental results.
In a recent analysis, new EDFs were calibrated to accommodate constraints from both CREX and PREX-2; however, this calibration shows differences from ab initio models and appears to conflict with the NICER observation due to the stiffening of the EOS at high densities relevant to neutron stars~\citep{Reed2024b}.

Alternatively, the nuclear electric dipole polarizability $\alpha_D$, which encodes the response of a nucleus to an external electric field, has emerged as a powerful probe of the symmetry energy~\citep{Reinhard2010}. The dipole polarizability imposes stringent constraints on the slope of the symmetry energy and, consequently, on the isovector properties of finite nuclei, such as neutron-skin thickness~\citep{Reinhard2010,Piekarewicz2012,ROCAMAZA201896}. Recent high-precision experiments on neutron-rich nuclei provide dipole polarizability in $^{48}$Ca~\citep{Birkhan2017}, $^{68}$Ni~\citep{PhysRevLett.111.242503}, $^{112,114,116,118,120,124}$Sn~\citep{PhysRevC.92.031305, Bassauer2020}, and $^{208}$Pb~\citep{PhysRevLett.107.062502}. These experimental results have been used to establish a link between dipole polarizability measurements and the symmetry energy in the EDFs, improving the EOS of neutron-rich matter~\citep{Piekarewicz2012,Rocamaza2013,Rokamaza2015,PhysRevC.90.064317,Zhang2015,Yuksel2019}.

Complementary to available constraints from nuclear experiments, which are mainly effective at low densities, astrophysical observations provide essential insights into the high-density regime of the EOS~\citep{Lattimer2021}. 
Key findings arise from measurements of neutron star masses, which can be precisely determined~\citep{Antoniadis-2013,Arzoumanian-2018,Cromartie-2020,Fonseca_2021,Romani-2022}, while radius remains significantly more challenging, even with the latest and more accurate constraints from NICER~\citep{Miller_2019,Miller_2021,Raaijmakers_2019,Riley_2019,Riley_2021,Dittmann_2024}. Recent analyses~\citep{Doroshenko-2022,Salmi_2024,Choudhury_2024} have also provided valuable information for the mass-radius plane.
The most stringent constraint on neutron star radii comes from the binary neutron star merger GW170817~\citep{PhysRevLett.119.161101,Abbott-2019}, where the gravitational-wave signal provided direct information on the tidal deformability of the merging stars, leading to improved radius estimates. The tidal deformability enables probing of the symmetry energy at $\sim$$2\rho_0$, which can be extrapolated toward saturation density using experimental data from finite nuclei~\citep{PhysRevLett.120.172702}. In this way, a direct connection between astrophysical and nuclear observables can be achieved.\\
\indent In this Letter, we introduce novel constraints on the neutron star EOS and its associated properties, based on recent experimental data on dipole polarizability in finite nuclei. We employ the relativistic density functional theory incorporating the density-dependent point-coupling (DD--PC) functionals~(\citealt{PhysRevC.78.034318}, \citeyear{NIKSIC20141808}), which provides a self-consistent description of both finite nuclei and neutron stars. 
Specifically, a family of $\beta$-equilibrated EOSs is considered, in which the symmetry energy at saturation density is systematically varied from $J=29$ to 36 MeV, with its corresponding slope ranging from $L=29$ to 94.1 MeV, thereby covering a wide range~\citep{Yuksel2021,KOLIOGIANNIS2025139362}. Although the symmetry energy spans a broad range, each DD--PC functional is calibrated to the same experimental data set and maintains a comparable level of accuracy for nuclear binding energies and charge radii around the stability line~\citep{Yuksel2019,PhysRevC.108.054305}. More details about the protocol employed to constrain the DD--PC family of functionals as well as the corresponding properties of finite nuclei and nuclear matter at saturation, are given in~\citet{Yuksel2021}.
The resulting EOSs are in good agreement with microscopic constraints from chiral effective field theory~\citep{PhysRevC.82.014314,Hebeler_2013,PhysRevC.103.025803}. 

\section{Methodology}
The symmetry energy $S_{2}(\rho)$ can be expanded around the saturation density $\rho_0$ in terms of a few bulk parameters as $S_{2}(\rho) = J + L u + \mathcal{O}(u^{2})$, where $u = (\rho - \rho_{0})/(3\rho_{0})$, and $J$ and $L$ represent the symmetry energy and its slope at saturation density, respectively~\citep{ROCAMAZA201896}. Around $\rho_0$, the values of $J$ and $L$ can be constrained through isovector nuclear observables such as neutron-skin thickness or dipole polarizability.
However, this expansion is not employed in the construction of the neutron star EOSs considered in this work. Instead, the EOSs are derived directly from microscopic nuclear matter calculations, as described in~\citet{KOLIOGIANNIS2025139362}. It is worth noting that while higher-order terms in the expansion, such as the curvature $K_{\rm sym}$ and skewness $Q_{\rm sym}$, may become increasingly relevant at suprasaturation densities, they remain poorly constrained by current experimental data~\citep{PhysRevLett.126.172503,Reed2024b}. Consequently, their uncertain contributions could introduce substantial model dependence in predictions of neutron star observables such as radii and tidal deformabilities. By relying on a microscopic treatment of the symmetry energy, rather than on a truncated density expansion, the present framework avoids such ambiguities and provides more robust predictions across the full density range relevant to neutron star interiors.

As a first step, we establish a direct connection between experimentally measured values of $\alpha_D$ across various nuclei and the slope parameter $L$. Then, by using the averaged value of $L$, we link $\alpha_D$ to astrophysical observables, such as the dimensionless tidal deformability $\Lambda_{1.4}$ and neutron star radius $R_{1.4}$ at $1.4~M_{\odot}$, thereby establishing a bridge between experiments on finite nuclei and neutron star structure.
Specifically, the electric dipole polarizability $\alpha_D$, which serves as a key observable for constraining the symmetry energy parameters, 
is defined through the inverse energy-weighted sum of the dipole response, primarily governed by the properties of isovector giant dipole resonance, a collective oscillation mode of neutrons against protons~\citep{ROCAMAZA201896}.
After determining the electric dipole strength $S(E1;E)$ as a function of the excitation energy $E$, the dipole polarizability $\alpha_D$ is calculated as
\begin{equation}
    \label{eq:alphaD}
    \alpha_{D}=\frac{8\pi e^{2}}{9}\int_{0}^{\infty}E^{-1}S(E1;E)dE,
\end{equation}
where the $E1$ strength function is evaluated using the
relativistic quasiparticle random phase approximation (RQRPA;~\citealt{PhysRevC.67.034312}), adopted for the point-coupling functionals (for a more detailed discussion, see \hyperref[appendix:dipole_polarizability]{Appendix}). Similarly, in an astrophysical context, tidal deformability characterizes the response of a neutron star to tidal forces exerted by a companion. It is calculated following the procedure described in~\citet{KOLIOGIANNIS2025139362}.

\begin{figure}[!t]
    \centering
    \includegraphics[width=\columnwidth]{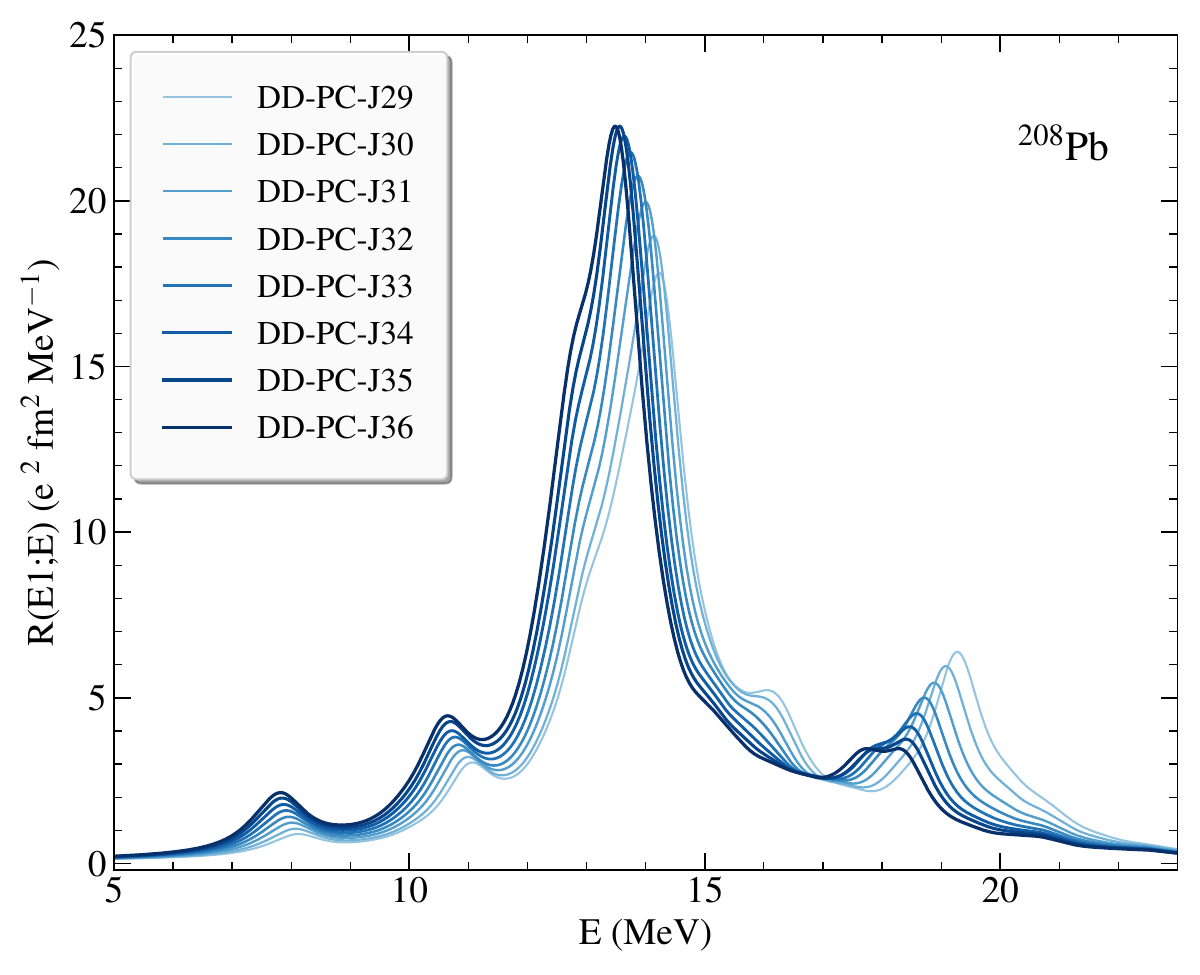}
    \caption{The electric dipole response $R(E1;E)$ of $^{208}$Pb as a function of the excitation energy $E$ for the DD--PC family of functionals. Darker colors indicate higher symmetry-energy values.}
    \label{fig:Pb208_E1}
\end{figure}

\section{Results}
As an illustrative example, we first consider $^{208}$Pb, a benchmark nucleus due to its doubly magic nature and substantial neutron excess. These properties make $^{208}$Pb an ideal reference nucleus for studying isovector properties in nuclear models.\\
\indent Figure~\ref{fig:Pb208_E1} displays the isovector electric dipole response $R(E1;E)$ of $^{208}$Pb (see also Equation~\eqref{eq:cont_response}), computed within the RQRPA using the DD--PC family of functionals in which the symmetry energy $J$ is systematically varied from 29 to 36~MeV at saturation density. The dipole response exhibits a pronounced sensitivity to the value of $J$. By using functionals with increasing $J$, the strength distribution progressively shifts toward lower excitation energies, reflecting a reduction of the isovector restoring force at the sub-saturation densities that govern the dynamics of the mode. This softening enhances the electric dipole polarizability $\alpha_D$ through the increased contribution of low-energy strength (see Equation~\eqref{eq:alphaD}). Thus, the controlled variation of $J$ within the DD--PC functional series provides a well-defined framework to explore how 
the symmetry energy shapes the global features of the dipole response and the dipole polarizability, with direct implications for the neutron-skin thickness and neutron star structure.

\begin{figure}[!t]
\centering
\includegraphics[width=\columnwidth]{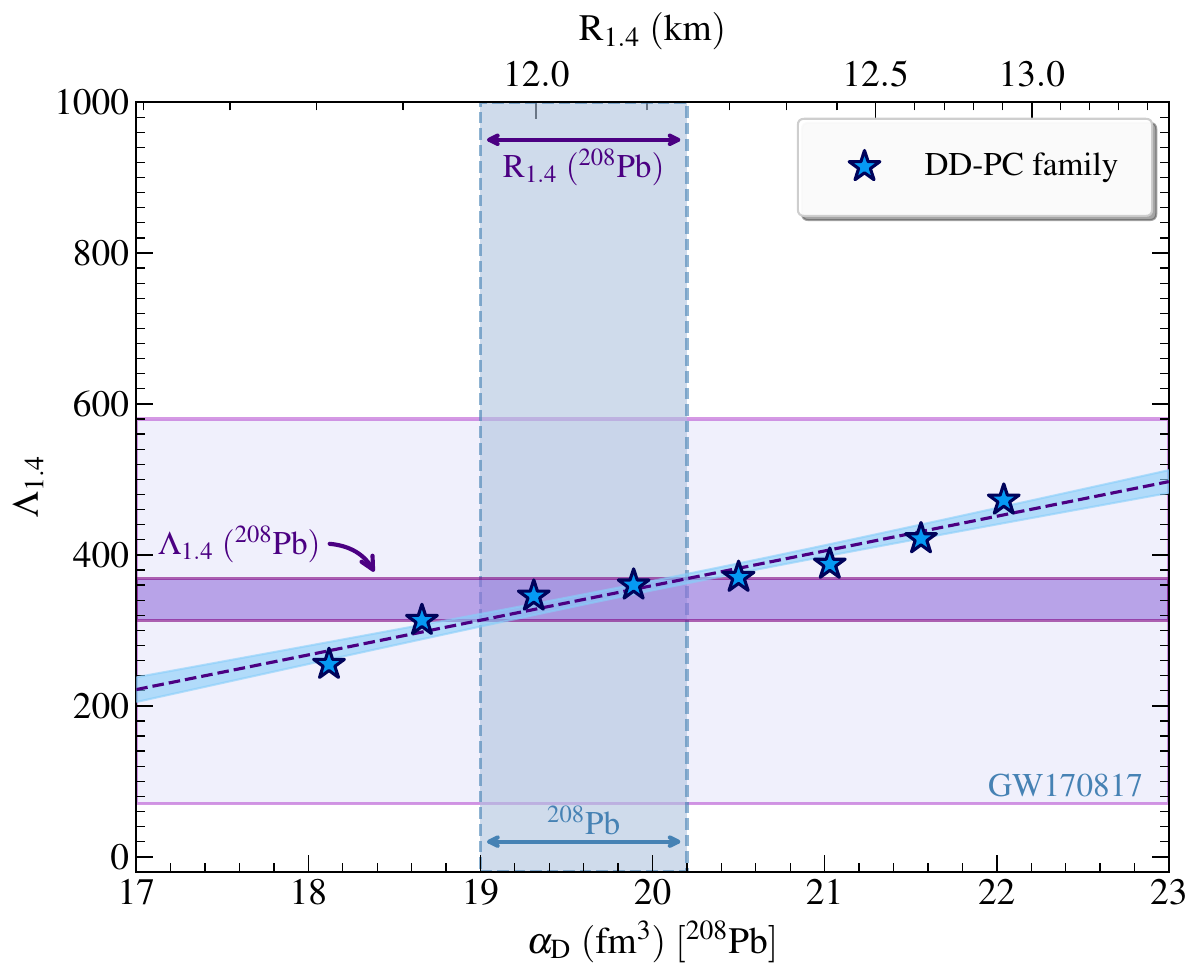}
\caption{The dimensionless tidal deformability of a $1.4~\rm{M_{\odot}}$ neutron star $\Lambda_{1.4}$ as a function of the electric dipole polarizability $\alpha_D$ of $^{208}$Pb and the corresponding neutron star radius for DD--PC EOSs. The horizontal shaded regions mark the limits derived from GW170817 event~\citep{Abbott-2019} and those obtained in the present study, while the vertical band denotes the experimental result for $\alpha_D[^{208}\text{Pb}]$ and the corresponding neutron star radius inferred in this work.}
\label{fig:L_aD_Pb}
\end{figure}

\begin{figure*}
    \centering
    \includegraphics[width=\textwidth]{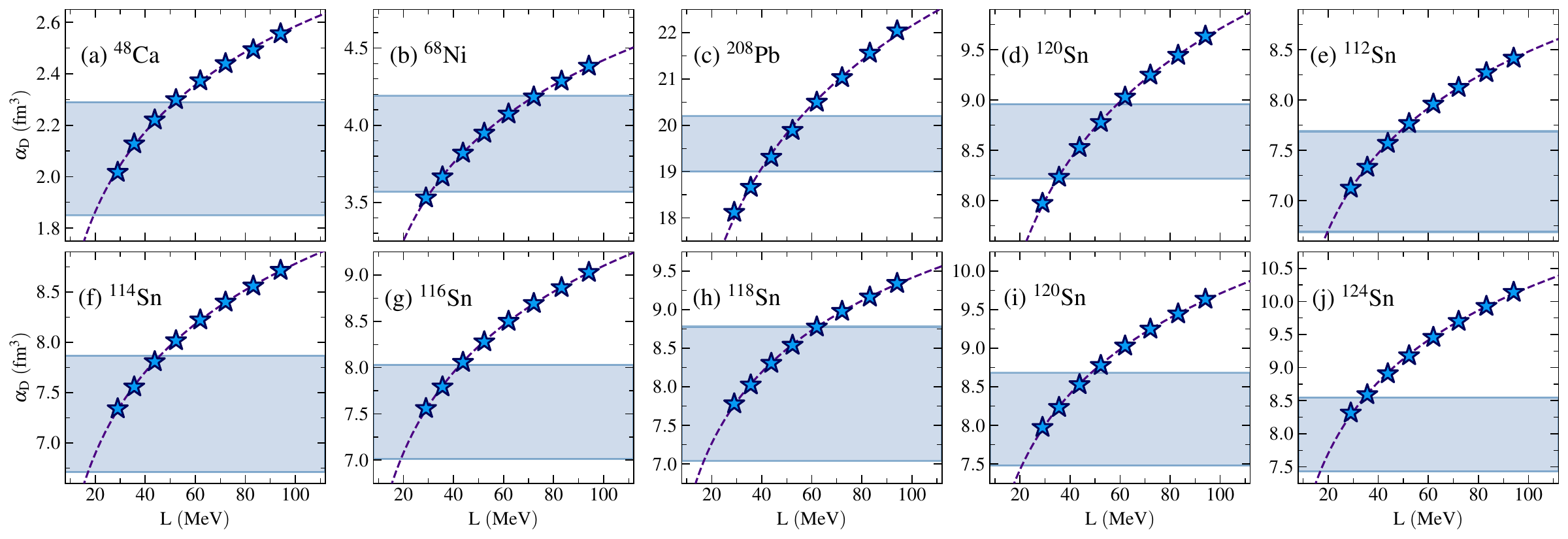}
    \caption{The electric dipole polarizability $\alpha_D$ as a function of the slope of the symmetry energy $L$ for the DD--PC family of functionals and various isotopes. The horizontal shaded regions represent experimental values on $\alpha_D$: (a) $^{48}$Ca~\citep{Birkhan2017}, (b) $^{68}$Ni~\citep{PhysRevLett.111.242503}, (c) $^{208}$Pb~\citep{PhysRevLett.107.062502,Rokamaza2015}, (d) $^{120}$Sn~\citep{PhysRevC.92.031305,Rokamaza2015}, and (e-j) $^{112-124}$Sn~\citep{Bassauer2020}.}
    \label{fig:L_ddpc}
\end{figure*}

Figure~\ref{fig:L_aD_Pb} presents predictions from the DD--PC family of EOSs for the dimensionless tidal deformability of a $1.4~M_{\odot}$ neutron star, $\Lambda_{1.4}$, as a function of both the electric dipole polarizability $\alpha_{D}$ and the corresponding neutron star radius $R_{1.4}$. The signal from GW170817~\citep{Abbott-2019}, which imposes an upper limit on the neutron star radius, $R_{1.4} = 13.366~{\rm km}$~\citep{KOLIOGIANNIS2025139362}, does not exclude any EOS from the DD--PC family due to the large uncertainty of the observation data. Since all considered EOSs satisfy this limit, the dipole polarizability constraint provides the most stringent bounds on both $\Lambda_{1.4}$ and $R_{1.4}$, as shown in Figure~\ref{fig:L_aD_Pb}. 
Specifically, the dipole polarizability of $^{208}$Pb yields refined ranges of $313.36\leqslant\Lambda_{1.4}\leqslant 368.47$, $11.96 \leqslant R_{1.4} \leqslant 12.15~{\rm km}$, and a corresponding symmetry energy slope of $39.12 \leqslant L \leqslant 55.93$ MeV. We note that the radius of a neutron star, particularly for intermediate-mass configurations around $1.4~M_{\odot}$, is influenced not only by the EOS of the homogeneous core but also by the inhomogeneous crust. While the outer crust is relatively well constrained and modeled using the Feynman-Metropolis-Teller~\citep{PhysRev.75.1561} and Baym, Pethick, and Sutherland~\citep{Baym-71} EOSs, the inner crust remains more uncertain due to its complex nuclear structure. In this work, we adopt the SLy EOS~\citep{Douchin_2001} for the inner crust, with the crust-core transition density determined self-consistently via the thermodynamical method for each EOS~\citep{Paar2014}.\\
\indent While a neutron-rich nucleus like $^{208}$Pb is highly relevant to neutron star matter, a comprehensive analysis, incorporating available experimental data on dipole polarizability across multiple neutron-rich nuclei, is expected to provide more consistent and 
stringent constraints on $L$, $\Lambda_{1.4}$, and $R_{1.4}$. In addition, analyzing multiple nuclei tests the robustness of this approach and further refines the density dependence of the symmetry energy. Figure~\ref{fig:L_ddpc} illustrates the relationship between the calculated dipole polarizability and the slope of the symmetry energy for the family of DD--PC functionals. The results are presented for various neutron-rich nuclei with complete experimental data on $\alpha_D$, including $^{48}$Ca, $^{68}$Ni, $^{112,114,116,118,120,124}$Sn, and $^{208}$Pb~\citep{PhysRevLett.107.062502,PhysRevLett.111.242503,PhysRevC.92.031305,Birkhan2017,Bassauer2020}.\\
\indent The consistent trends exhibited by DD--PC functionals across these nuclei, combined with their experimental dipole polarizability values, impose a set of independent constraints on the slope of the symmetry energy. To estimate $L$ from the measured $\alpha_D$, two cases are considered: (a) a subset of 4 nuclei, denoted as CNSP-4 [$^{48}$Ca~\citep{Birkhan2017}, $^{68}$Ni~\citep{PhysRevLett.111.242503}, $^{120}$Sn~\citep{PhysRevC.92.031305}, $^{208}$Pb~\citep{PhysRevLett.107.062502}],  
and (b) the full set of 10 nuclei, denoted as CNSP-10, supplementing the CNSP-4 nuclei with more recent measurements of even-even isotopes $^{112-124}$Sn~\citep{Bassauer2020}. The analysis is performed separately for the two cases because more recent measurements of dipole polarizability in Sn isotopes suggest that a relatively low symmetry energy is needed to reproduce the experimental data~\citep{Bassauer2020}.
In fact, two recent measurements of $^{120}$Sn~\citep{PhysRevC.92.031305,Bassauer2020} yield significantly different values of $\alpha_D$.
Through a direct comparison of these data, we conclude that this discrepancy originates from differences in the treatment of excitations above $\sim$20 MeV. In~\citet{Bassauer2020} these excitations are described by quasiparticle phonon model, while~\citet{PhysRevC.92.031305} incorporates measured data from~\citet{Fultz1969}. It is important to note that in the high-energy region above 30 MeV, the experimental dipole strength may contain a nonnegligible amount of contamination from nonresonant processes, known as the quasideuteron effect~\citep{Rokamaza2015}. Therefore, for a direct comparison with theoretical (Q)RPA calculations, these contributions must be removed, as was done for $^{120}$Sn and $^{208}$Pb in~\citet{Rokamaza2015}.\\
\indent In this work, we determine the weighted average of the slope of the symmetry energy for the two sets of nuclei $L = 46.08\pm 6.06~{\rm MeV}$ (CNSP-4) and $L = 37.16\pm 3.96~{\rm MeV}$ (CNSP-10). Using the same procedure, we also obtain the weighted average of the symmetry energy at saturation density $J = 31.34 \pm 0.79~{\rm MeV}$ (CNSP-4) and $J = 30.37\pm 0.64~{\rm MeV}$ (CNSP-10). By incorporating data from multiple nuclei, this approach reduces uncertainties and enhances the reliability of the inferred constraints on the density dependence of the symmetry energy.\\
\begin{figure}[t!]
\centering
\includegraphics[width=\columnwidth]{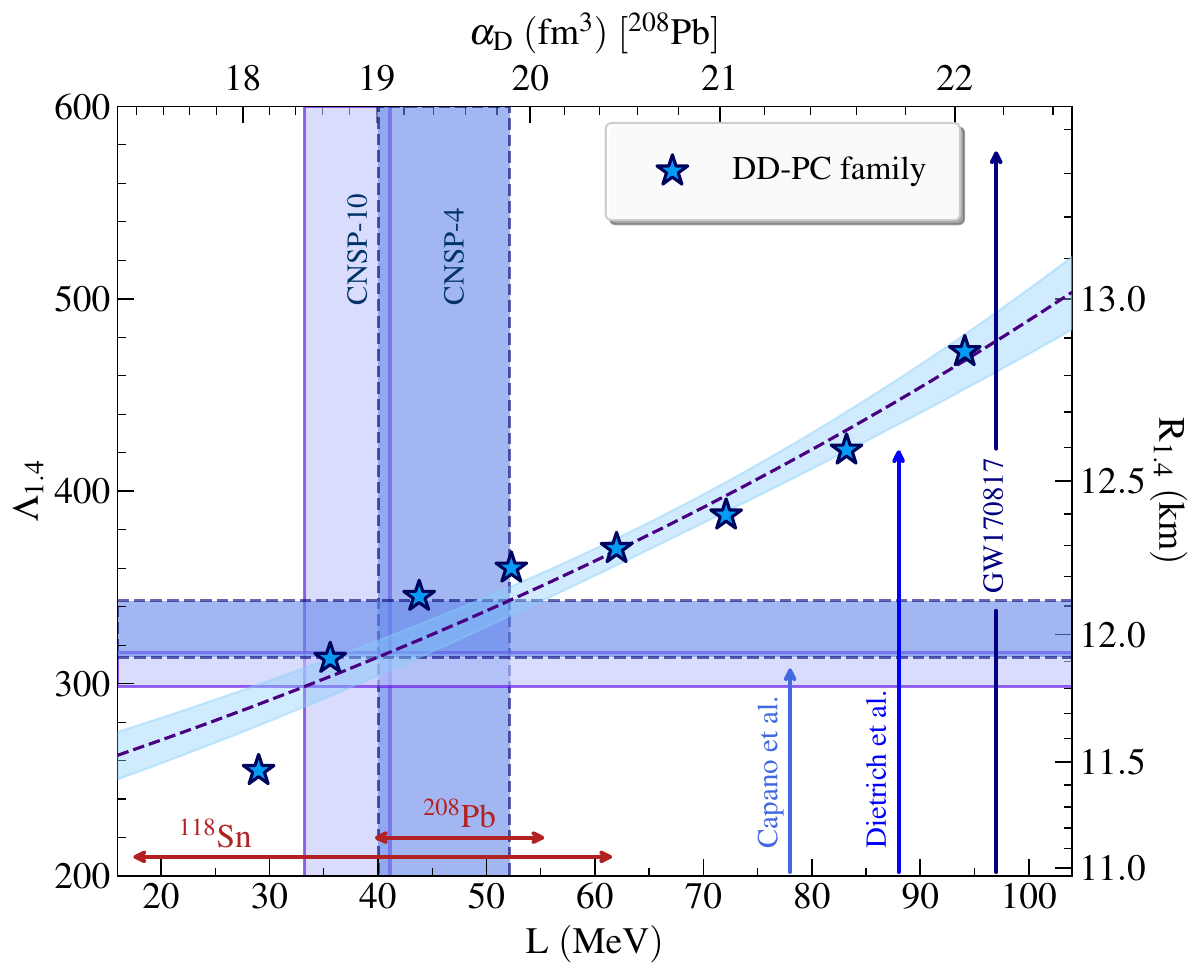}
\caption{The dimensionless tidal deformability of a $1.4~\rm{M_{\odot}}$ neutron star $\Lambda_{1.4}$ (left axis) and its corresponding radius $R_{1.4}$ (right axis) as a function of the slope of the symmetry energy $L$ for DD--PC EOSs. The dipole polarizability of $^{208}$Pb is also shown on the top axis for reference. The vertical shaded regions represent the $L$ values obtained from dipole polarizability for the CNSP-4 and CNSP-10 sets of nuclei, while the horizontal shaded regions denote the corresponding constraints for the $\Lambda_{1.4}~(\text{or}~R_{1.4})$, respectively. Vertical arrows indicate neutron star radii from~\citet{Capano-2020},~\citet{doi:10.1126/science.abb4317}, and $\Lambda_{1.4}$ from the GW170817 event~\citep{Abbott-2019}. Horizontal arrows mark nuclei with the largest and smallest deviations of $L$ value.}
\label{fig:L_R_L}
\end{figure}
\indent Figure~\ref{fig:L_R_L} shows the dimensionless tidal deformability of a $1.4~M_{\odot}$ neutron star as a function of the slope of the symmetry 
energy, calculated using the EOSs from the DD--PC family of functionals. The bands corresponding to $L$ values constrained by dipole polarizability data, CNSP-4 and CNSP-10, are also shown. Since the CNSP-4 case leads to slightly stiffer EOSs compared to CNSP-10 ($\Delta L \sim 10~{\rm MeV}$), the resulting $\Lambda_{1.4}$ values are also higher. Moreover, because the dimensionless tidal deformability scales with the neutron star radius in a power-law relation~\citep{KOLIOGIANNIS2025139362}, a similar trend is observed for the radius, which also increases with $\Lambda_{1.4}$. This dependence is depicted in Figure~\ref{fig:L_R_L}, where the right axis represents the corresponding neutron star radius. Specifically, for the two cases, the following ranges are obtained:
\[
\begin{array}{ll}
    \text{CNSP-4:} &
    \begin{cases}
        313.80 \leqslant \Lambda_{1.4} \leqslant 343.18, \\
        11.92 \leqslant R_{1.4} \leqslant 12.12~{\rm km},
    \end{cases} \\
    \text{CNSP-10:} &
    \begin{cases}
        298.39 \leqslant \Lambda_{1.4} \leqslant 316.35, \\
        11.81 \leqslant R_{1.4} \leqslant 11.94~{\rm km}.
    \end{cases}
\end{array}
\]
In addition, the same figure also denotes neutron star radii derived from neutron star merger analyses (vertical arrows;~\citealt{Capano-2020,doi:10.1126/science.abb4317}), along with the largest ($^{118}$Sn) and smallest ($^{208}$Pb) deviations of the $L$ value (horizontal arrows) for reference.
It should be noted that the constraints for $L$, $\Lambda_{1.4}$, and $R_{1.4}$ overlap in a rather narrow region, with CNSP-4 yielding higher values than CNSP-10. This outcome is a direct consequence of the inclusion of the dipole polarizability for Sn isotopes from~\citet{Bassauer2020} in the CNSP-10 set of nuclei. In that study, as previously discussed, measurements of $\alpha_D$ result in low values, and, in turn, in lower $L$ values. Since the $\alpha_D$ values for even-even isotopes $^{112-124}$Sn reported in~\citet{Bassauer2020} are not solely based on measured data but also include the higher-energy part of dipole transition strength above 20 MeV calculated within the quasiparticle phonon model, we consider our CNSP-4 results as more reliable for the neutron star properties. Clearly, further and more complete experimental studies are required to provide more consistent and model-independent results for dipole polarizability in Sn isotopes.

\begin{figure}[t!]
\centering
\includegraphics[width=\columnwidth]{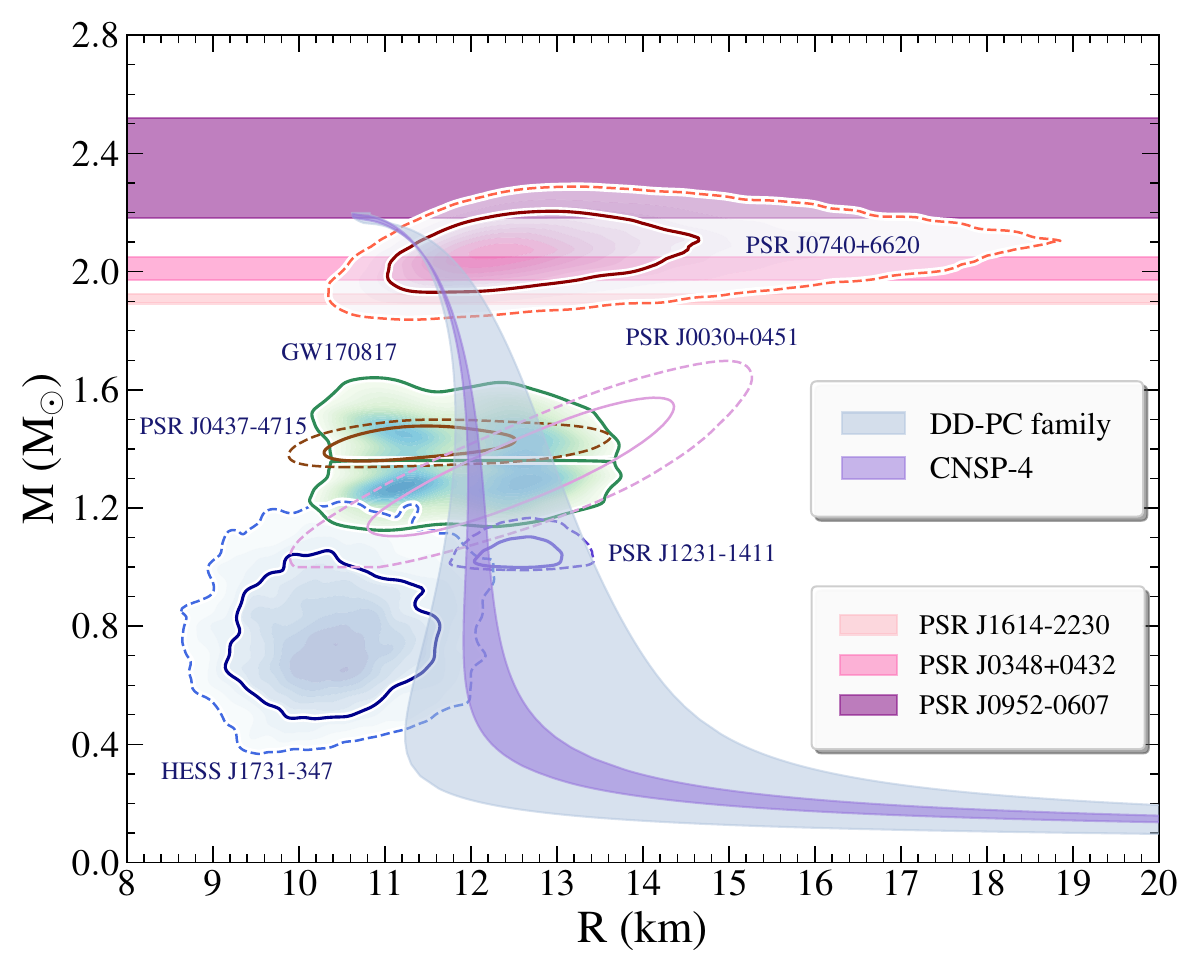}
\caption{Gravitational mass as a function of the radius for the DD--PC EOSs, denoted by the extended shaded region. The inner shaded region corresponds to the limits from the CNSP-4 set of nuclei. The shaded contours represent multiple observations, including HESS J1731-347~\citep{Doroshenko-2022}, PSR J1231-1411~\citep{Salmi_2024}, PSR J0030+0451~\citep{Raaijmakers_2019,Riley_2019}, PSR J0437-4715~\citep{Choudhury_2024}, the GW170817 event~\citep{Abbott-2019}, and PSR J0740+6620~\citep{Fonseca_2021,Dittmann_2024}, as well as maximum neutron star masses inferred from pulsar observations~\citep{Antoniadis-2013,Arzoumanian-2018,Romani-2022}.}
\label{fig:mr_plot}
\end{figure}

Both CNSP-4 and CNSP-10 indicate that the EOS is relatively soft in the vicinity of saturation density, consistent with the small neutron star radii inferred from gravitational wave signals of binary neutron star mergers~\citep{Abbott-2019}. 
This agreement between constraints from nuclear experiments and astrophysical observations suggests that the EOS remains soft up to densities corresponding to $1.4~M_{\odot}$ $(\sim 3\rho_{0})$, with no clear indication of a phase transition within this range. If a transition to exotic degrees of freedom were to occur at lower densities, a more pronounced softening would be expected, potentially conflicting with observed neutron star properties.

We conclude our analysis with Figure~\ref{fig:mr_plot}, which presents the neutron star gravitational mass as a function of radius for the EOSs of DD--PC functionals, depicted by the extended shaded region. 
This final figure consolidates constraints from various multimessenger observations, including HESS J1731-347~\citep{Doroshenko-2022}, NICER observations of PSR J1231-1411~\citep{Salmi_2024}, PSR J0030+0451~\citep{Miller_2019,Raaijmakers_2019,Riley_2019}, PSR J0437-4715~\citep{Choudhury_2024}, and PSR J0740+6620~\citep{Fonseca_2021,Miller_2021,Riley_2021,Dittmann_2024}, the GW170817 event~\citep{Abbott-2019}, as well as maximum neutron star masses inferred from pulsar observations~\citep{Antoniadis-2013,Arzoumanian-2018,Cromartie-2020,Fonseca_2021,Romani-2022}. Incorporating the more reliable data set for dipole polarizability, CNSP-4 refines the DD--PC EOS predictions, narrowing the extended mass-radius region to the inner shaded area. The consideration of dipole polarizability significantly reduces uncertainties in the neutron star radius while ensuring consistency with the predicted range~\citep{PhysRevLett.120.172703,PhysRevLett.120.172702,PhysRevC.98.035804,TSANG20191,KOLIOGIANNIS2025139362}.
Notably, this inner region not only aligns with nuclear experimental constraints but also remains consistent with all considered astrophysical observations, including those from NICER. This highlights the strong agreement between nuclear physics constraints and astrophysical observations, underscoring the robustness of the current neutron-star models.

\section{Conclusion}
The presented results demonstrate that dipole polarizability serves as a valuable constraint on neutron-star properties through the slope of the symmetry energy, a key quantity governing the nuclear matter EOS. Dipole polarizability in nuclei emerges as a promising alternative to the neutron-skin thickness, which faces large uncertainties and poses challenges for nuclear models attempting to reproduce both the $^{48}\mathrm{Ca}$ and $^{208}\mathrm{Pb}$ results from recent parity-violating electron-scattering experiments~\citep{Adhikari2021,Adhikari2022}. Although the CREX and PREX-2 measurements are individually consistent within their quoted uncertainties, their simultaneous description within a single theoretical framework remains difficult. This tension highlights the significance of $\alpha_D$ as an alternative probe of the density dependence of the symmetry energy, that is accessible in high-precision experiments.

In this context, advancing our understanding of dipole polarizability will significantly improve predictions of neutron-star properties and reduce uncertainties in theoretical models. Future measurements of dipole polarizability across various neutron-rich nuclei will be particularly important, as averaging the slope of the symmetry energy from independent experiments will further reduce uncertainties in related neutron-star properties. As highlighted in this study, expanding experimental investigations of dipole transitions beyond the current limit of $\sim 20$ MeV, especially for even-even isotopes $^{112-124}$Sn~\citep{Bassauer2020}, is essential, given their sizable contribution to the total dipole polarizability.

In addition, the signal from the binary neutron star merger GW170817 has provided significant insights into the nature of dense nuclear matter. Combined with NICER pulsar observations constraining the mass and radius of neutron stars, these multimessenger observations have imposed key constraints on the nuclear matter EOS. As such observations remain our primary window into the properties of dense matter, the forthcoming Einstein Telescope, with its enhanced sensitivity in the lower frequency band, in collaboration with LIGO, Virgo, and KAGRA, is expected to further advance our understanding by detecting fainter signals and extending the reach of neutron star observations.

%% Please use the acknowledgment and contribution environments. This will 
%% be anonomyized when the "anonymous" style option is used. 
\begin{acknowledgments}
We are indebted to P. von Neumann-Cosel, A. Tamii, and I. Fri{\v s}{\v c}i{\'c} for useful discussions which helped to improve various aspects of this study.
This work is supported by the Croatian Science Foundation under the project number HRZZ-MOBDOL-12-2023-6026 and under the project Relativistic Nuclear Many-Body Theory in the Multimessenger Observation Era (HRZZ-IP-2022-10-7773). E.Y. acknowledges support from the UK STFC under award no. ST/Y000358/1.
N.P. acknowledges support from the project “Implementation of cutting-edge research and its application as part of the Scientific Center of Excellence for Quantum and Complex Systems, and Representations of Lie Algebras”, Grant No. PK.1.1.10.0004, co-financed by the European Union through the European Regional Development Fund - Competitiveness and Cohesion Programme 2021-2027.
\end{acknowledgments}

\begin{contribution}
The conceptual framework of the study was developed by P.S.K. and N.P. Methodological design and computational implementation were carried out by P.S.K. and N.P., with P.S.K. responsible for model validation, formal analysis, visualization, and figure preparation. Model calculations and data collection and curation were performed by P.S.K., E.Y., and T.G. The manuscript was drafted by P.S.K. and N.P., and revised by all authors. Supervision and project coordination were provided by N.P. Financial support for this work was secured by P.S.K., E.Y., and N.P.
\end{contribution}

\appendix

\section{Dipole Polarizability and QRPA Response Function}
\label{appendix:dipole_polarizability}

In the quasiparticle random-phase approximation (QRPA), nuclear excited states $|\nu\rangle$ are described as small-amplitude oscillations of the Hartree--Bogoliubov ground state $|0\rangle$. The QRPA eigenvalue problem to solve is given by~\citet{ring2004nuclear,PhysRevC.67.034312,colo2013self}
\begin{equation}
    \begin{pmatrix}
    A & B \\
    -B^* & -A^*
    \end{pmatrix}
    \begin{pmatrix}
    X^\nu \\ Y^\nu
    \end{pmatrix}
    =
    E_\nu
    \begin{pmatrix}
    1 & 0 \\ 0 & -1
    \end{pmatrix}
    \begin{pmatrix}
    X^\nu \\ Y^\nu
    \end{pmatrix},
\end{equation}
where $E_\nu$ denotes the excitation energy of the $\nu$-th QRPA mode, and $X^\nu$, $Y^\nu$ are the forward and backward amplitudes, respectively, normalized as $\sum_{kk'}(|X_{kk'}^\nu|^2 - |Y_{kk'}^\nu|^2)=1$. In the QRPA framework, the matrices $A$ and $B$ encode the residual two-body interaction between quasiparticles, including contributions from the particle–hole and particle–particle channels. They are defined in the two--quasiparticle basis $\{|kk'\rangle\}$ as
\begin{align}
    A_{kk'll'} &= (E_k + E_{k'})\,\delta_{kl}\delta_{k'l'}
    + \langle kk'|\widetilde{V}_{\mathrm{res}}|ll'\rangle, \nonumber \\
    B_{kk'll'} &= \langle kk'|\widetilde{V}_{\mathrm{res}}|l'l\rangle,
\end{align}
where $E_k$ are the quasiparticle energies obtained from the Hartree–Bogoliubov calculation, and $\widetilde{V}_{\mathrm{res}}$ is the antisymmetrized residual interaction derived self-consistently from the underlying energy--density functional (in this work, the DD--PC family). The diagonal term $(E_k+E_{k'})$ represents the unperturbed energy of a two--quasiparticle configuration, while the residual-interaction matrix elements $\langle kk'|\widetilde{V}_{\mathrm{res}}|ll'\rangle$ couple different configurations.

The excitation properties of each QRPA mode are analyzed through external one-body operators that connect the ground state and excited configurations. In particular, the isovector electric dipole operator is defined as
\begin{equation}
    \hat{Q}_{1M}^{(\mathrm{IV})}
    = \frac{N}{A}\sum_{p=1}^{Z} r_p\,Y_{1M}(\hat{r}_p)
    - \frac{Z}{A}\sum_{n=1}^{N} r_n\,Y_{1M}(\hat{r}_n),
\end{equation}
where $N$, $Z$, and $A$ denote the neutron, proton, and mass numbers, respectively; $r_{n(p)}$ represents the radial coordinate of each neutron (proton); and $Y_{1M}(\hat{r})$ is the spherical harmonic. Moreover, the transition strength for the $E1$ mode is
\begin{equation}
    S(E1; 0 \!\to\! \nu)
    =
    \left|
    \sum_{kk'}
    \bigl( X_{kk'}^{\nu} + Y_{kk'}^{\nu} \bigr)
    \langle k \| \hat{Q}_{1}^{(\mathrm{IV})} \| k' \rangle
    \right|^{2},
\end{equation}
where $\langle k\|\hat{Q}_{1}^{(\mathrm{IV})}\|k'\rangle$ denotes the corresponding reduced matrix element of the isovector dipole operator. 
The discrete transition probabilities $S(E1;0\!\to\!\nu)$ define the dipole strength function
\begin{equation}
    S(E1;E) = \sum_\nu S(E1;0\!\to\!\nu)\,\delta(E - E_\nu),
\end{equation}
which represents the $E1$ response of the nucleus as a function of the excitation energy~$E$.

The electric dipole polarizability $\alpha_D$ quantifies the linear response of a nucleus to an external static electric field and serves as a global measure of the electric dipole ($E1$) strength distribution. It can be obtained as the inverse energy--weighted moment of the dipole response function \citep{BOHIGAS1981105,LIPPARINI1989103,Reinhard2010}:
\begin{equation}
    \alpha_{D} = \frac{8\pi e^{2}}{9} \int_{0}^{\infty} \! E^{-1} S(E1;E)\ dE.
\end{equation}
The dipole polarizability is particularly sensitive to the neutron--proton density distribution and, consequently, to the isovector properties of the nuclear energy--density functional, including the symmetry energy and its density dependence.

After performing the QRPA calculations and obtaining the discrete dipole strengths $S(E1;0\!\to\!\nu)$, the spectrum is folded with a Lorentzian of width $\Gamma = 1.0$~MeV. The resulting continuous response is
\begin{equation}
R(E1;E)
=
\sum_{\nu}
\frac{1}{2\pi}
\frac{\Gamma}{(E-E_\nu)^{2} + \Gamma^{2}/4}\,
  S(E1; 0 \!\to\! \nu),
\label{eq:cont_response}
\end{equation}
yielding a smooth distribution suitable for studying the giant dipole resonance.

\bibliography{bibliography}{}
\bibliographystyle{aasjournalv7}


\end{document}